\documentclass[superscriptaddress,super,sort&compress,numbers,a4paper,10pt,twocolumn,amssymb,amsmath,aps,prl]{revtex4-1}

\usepackage[margin=1.8cm]{geometry}

\providecommand{\sspace}{\hspace*{0.15em}}

\providecommand{\dfigwidth}{\textwidth}
\providecommand{\sfigwidth}{\columnwidth}

\setcitestyle{super,open={},close={}}

\usepackage{graphicx}
\usepackage{dsfont}
\usepackage{xfrac}
\usepackage[usenames,dvipsnames]{xcolor}
\usepackage[pdftex]{hyperref}

\providecommand{\ket}[1]{\ensuremath{| #1\rangle}}

\providecommand{\ketbra}[2]{\ensuremath{| #1\rangle \! \langle #2 |}}
\providecommand{\matelm}[3]{\ensuremath{\langle #1 | #2 | #3 \rangle}}

\providecommand{\negation}[1]{ \neg {#1} }

\providecommand{\addedrichard}[1]{#1}
\providecommand{\johnadded}[1]{#1}
\providecommand{\ronaldremoved}[1]{}
\providecommand{\replymooji}[1]{#1}

\providecommand{\OxfordMat}{University of Oxford, Department of Materials, 12/13 Parks Road, Oxford, OX1 3PH, United Kingdom}
\providecommand{\OxfordPhil}{University of Oxford, Faculty of Philosophy, 10 Merton Street, Oxford, OX1 4JJ, United Kingdom}
\providecommand{\Delft}{Kavli Institute of Nanoscience Delft, Delft University of Technology, Post Office Box 5046, 2600 GA Delft, The Netherlands}
\providecommand{\ElementSix}{Element Six, Ltd., Kings Ride Park, Ascot, Berkshire SL5 8BP, United Kingdom}

\providecommand{\SuppMat}{Supplementary online information} 

\begin{document}

\title{Opening up the Quantum Three-Box Problem with Undetectable Measurements}

\author{Richard E. George}
\email{richard.george@materials.ox.ac.uk}
\affiliation{\OxfordMat}
\author{Lucio Robledo}
\email{l.m.robledoesparza@tudelft.nl}
\affiliation{\Delft}
\author{Owen Maroney}
\affiliation{\OxfordPhil}
\author{Machiel Blok}
\author{Hannes Bernien}
\affiliation{\Delft}
\author{Matthew L. Markham}
\author{Daniel J. Twitchen}
\affiliation{\ElementSix}
\author{John J. L. Morton}
\author{G. Andrew D. Briggs}
\affiliation{\OxfordMat}
\author{Ronald Hanson}
\affiliation{\Delft}

\date{\today}
 

\begin{abstract}
One of the most striking features of quantum mechanics is the profound effect exerted by measurements alone~\cite{Julsgaard:2001by,Knee:2011uva,Bernien:2012hy}.
Sophisticated quantum control is now available in several experimental systems~\cite{Gao:2010cm,DiCarlo:2010js,Burrell:2010kh},
exposing discrepancies between quantum and classical mechanics whenever measurement induces disturbance of the interrogated system. In practice, such discrepancies may frequently be explained as the back-action required by quantum mechanics adding quantum noise to a classical signal~\cite{Clerk:2010dh}. 
Here we implement the `three-box' quantum game of Aharonov and Vaidman~\cite{AHARONOV:1991ux} in which quantum measurements add no detectable noise to a classical signal, by utilising state-of-the-art control and measurement of the nitrogen vacancy centre in diamond~\cite{Robledo:2011fs,Bernien:2012hy}.
Quantum and classical mechanics then make contradictory predictions for the same experimental procedure, however classical observers cannot invoke measurement-induced disturbance to explain this discrepancy. We quantify the residual disturbance of our measurements and obtain data that rule out any classical model by $\gtrsim 7.8$ standard deviations, allowing us for the first time to exclude the property of macroscopic state-definiteness from our system. Our experiment is then equivalent to a Kochen-Spekker test of quantum non-contextuality~\cite{Kochen:1967vo} that successfully addresses the measurement detectability loophole~\cite{Leifer:2005ek}.
\end{abstract}

\maketitle

Classical physics describes the nature of systems that are `large' enough to be considered as occupying one definite state in an available state space at any given time$^{12}$.
~Macrorealism (MR) then applies whenever it is possible to perform non-disturbing measurements to identify this state without significantly modifying the system's subsequent behaviour~\cite{LEGGETT:1985td}. Macrorealism allows the assignment of a definite history (or probabilities over histories) to classical systems of interest, but the MR condition can break down for systems `small' enough to be quantum mechanical (QM) during times `short' enough to be quantum coherent  --- times and distances that now exceed seconds~\cite{Tyryshkin:2011fi} and millimeters~\cite{DiCarlo:2010js} in the solid state. How can we tell whether a particular case is better described by QM or MR? If there is a cross-over between these, what does it represent? 

One explanation for the breakdown of MR is that measurement back-action (either deliberate measurements by an experimenter or effective measurements from the environment) unavoidably change the state in the quantum limit, excluding macrorealism due to a breakdown of non-disturbing measurability. This position is supported by `weak value' experiments~\cite{Goggin:2011iw,Massar:2011jf} that explore the transition from quantum to classical behaviour as a measurement coupling is varied; Quantum behaviour is found under weak coupling, whilst MR-compatible behaviour is recovered when strong projective measurements effectively `impose' a classical value onto the measured quantum system~\cite{Goggin:2011iw}.

We examine a case in which the back-actions of sequential `strong' projective measurements impose new quantum states that provide no detectable indication of disturbance to a `macrorealist' observer. We show these state are {\it still} incompatible with macrorealism, however, as no possible MR-compatible history can be assigned to the process as a whole. Our experiment can be described as a game played by two opponents (Alice and Bob) who take alternate turns to measure a shared system. The system they share may or may not obey the axioms of macrorealism.  For the purposes of the game, Bob assumes he may rely on the MR assumptions being true, and only Alice is permitted to manipulate the system between measurements. If Bob is correct to assume MR holds, the game they play is constructed in his favour, yet `paradoxically' the exact same sequence of operations will define a game that favours Alice when a quantum-coherent description of the system is valid~\cite{Aharon:2008fv}. 

\begin{figure*}[t]
\begin{center}
\includegraphics[width=\dfigwidth]{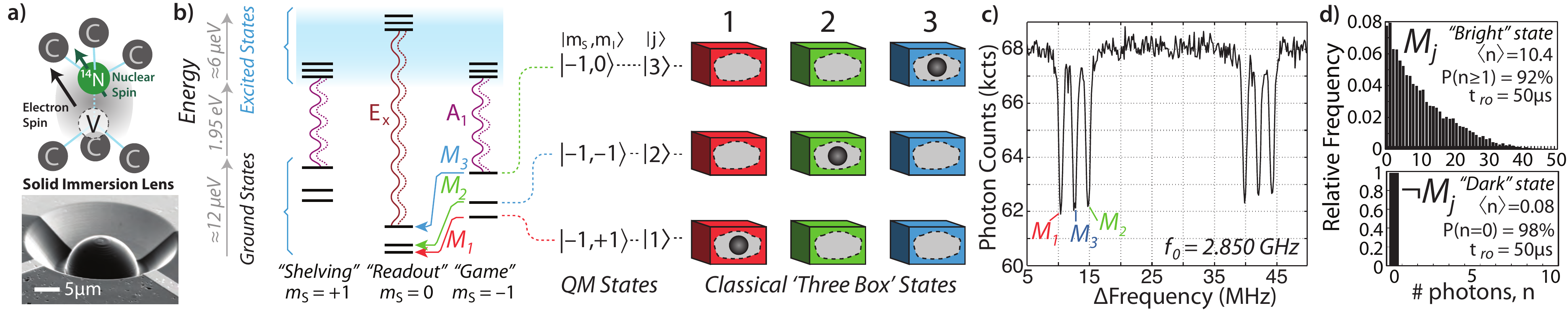}
\end{center}
\caption{ \label{fig1}
{\bf The `three-box' game is implemented using the $^{14}$N nuclear spin of the NV$^-$ centre in diamond, measured using the electron spin.
a)} Schematic of the NV$^-$ defect in diamond and representative diamond lens used in the measurements.
{\bf b)} The magnetic moment of the electron spin is quantised into one of three values, $m_S = -1,0,+1$. These states split into a further three ($m_I = -1,0,+1$) according to magnetic moment of the $^{14}$N nuclear spin. The $m_S = \pm 1$ states fluoresce via the A$_1$ transition, whilst $m_S = 0$ fluoresces via the E$_x$ transition.  We use the $m_S = -1$ manifold to hold the three states in the game, conditionally moving the state between $m_S = -1$ and $m_S = 0$ dependent on the nuclear spin sublevel $m_I$. These three $m_I$ states are taken to correspond to the configurations of a hidden ball.
{\bf c)} We identify the allowed microwave transitions $( \Delta m_S = 1, \Delta m_I = 0 )$, that provide the $M_j$ readouts.
{\bf d)} Photon counting statistics, in each case from 10,000 trials, observed during a typical projective read-out indicate the presence (top) or absence (bottom) of optical fluorescence, corresponding to outcomes $M_j$ and $\negation{M}_j$ respectively.
} 
\end{figure*}

Experimentally, we use the $^{14}$NV$^-$ centre ($S\!=\!1$, $I\!=\!1$) in diamond as Alice and Bob's shared system, enabling us for the first time to maintain near-perfect undetectability of Bob's observations. The experiment involves a three level pre- and post-selection~\cite{AHARONOV:1964ul,Massar:2011jf} that is known to be equivalent to a Kochen-Spekker test of quantum non-contextuality~\cite{Leifer:2005ek}. Such tests are only possible in $d \! \ge \! 3$ Hilbert spaces~\cite{Kochen:1967vo}, and recent advances in the engineering~\cite{Bernien:2012hy} and control~\cite{Robledo:2011fs} of the NV system enable the multiple projective non-demolition measurements that are crucial to observing Alice's quantum advantage in the lab. We describe the game~\cite{AHARONOV:1991ux} and Bob's verification of it from the MR perspective, then discuss the experiment and results from the QM position. We quantify the incompatibility of our results with macrorealism through use of a Leggett-Garg inequality~\cite{LEGGETT:1985td} and discuss the implications of our result. 


In the `three-box' quantum game~\cite{AHARONOV:1991ux}, Alice and Bob each inspect a freshly prepared three state system (classically, three separate boxes, hiding one ball) using an apparatus that answers the question: ``Is the system now in state $j$?'' (``Is the ball in box $j$?'') for $j=1,2,3$ by responding either true ($1$) or false ($0$). The question is answered by performing one of three mutually orthogonal measurements $M_j$.  The game allows Bob a single use of either $M_1$ or $M_2$. Alice is allowed only to use $M_3$, and additionally to manipulate the system. Alice is allowed one turn (a manipulation either before or after an $M_3$ measurement) before Bob to prepare the system, and one turn following him. Alice attempts to guess Bob's measurement result, and the pair bet on Alice correctly answering: ``Did Bob find $M_j$ to be true?''. Alice offers Bob $\gg 50\%$ odds to predict when his $M_j$ was true, although she may `pass' on any given round at no cost when she is uncertain.

Bob realises that if the $M_j$ measurements are performed on a system following MR axioms, Alice must bet incorrectly $\ge \! 50\%$ of the time, even if Alice could `cheat' by knowing which $j$-value will be presented (classically, knowing which box contains the ball); yet with three boxes and his free choice between $M_1$ or $M_2$, Alice is prevented from using such knowledge to win with $> 50\%$ success rate.  Bob expects to win if the $M_j$ measurements reproduce the behaviour of a ball hidden in one of three boxes. The conditions for this are: {\bf a)} The $M_j$ measurements are repeatable and mutually exclusive, so that $ M_j \wedge M_k = \delta_{jk}$; (classically, the ball does not move when measured); {\bf b)} for any trial $M_1 \vee M_2 \vee M_3 =1$ (there is only one ball, it is definitely in one of the boxes); and  {\bf c)} Bob has equal probability to find each $j$-value when measuring a fresh state, with $P_{M_j}(B)=1/3 \; j \in 1,2,3$. (The ball is placed at random). The conditions {\bf a)}~-~{\bf c)} serve to prevent Alice from learning Bob's $M_j$ result in any macroreal system. \replymooji{Before accepting Alice's invitation to play, Bob verifies properties {\bf a)} - {\bf c)} hold experimentally, by carrying out $M_j$ measurements. During verification, the game rules are relaxed and Bob is permitted to make pairs of sequential measurements, checking $ M_j \vee M_k = \delta_{jk}$; He is also allowed to measure every $M_j$ including $M_3$, which will be reserved for Alice once betting commences.

When Bob is satisfied that {\bf a) - c)} hold, the game appears fair from his macrorealist standpoint. Bob accepts Alice's wager and play commences with Alice preparing a state, which Bob measures using either $M_1$ or $M_2$, whilst keeping his $j$-choice and $M_j$-result secret.} Alice manipulates the system, uses her $M_3$ measurement and bets whenever her $M_3$-result is true. Believing that Alice could only guess his result, Bob accepts Alice's wager. Doing so, he finds Alice's probability of obtaining a true $M_3$-result is $P_{M_3}(A) \simeq 1/9$, independent of his $j$-choice. Under macrorealism, Bob could account for this only through Alice using a non-deterministic manipulation that would reduce the information available to her from the $M_3$-result. \replymooji{To Bob's surprise, when Alice plays, her $M_3$-true results coincide with rounds on which Bob's $M_j$-result was also true. She passes whenever Bob's $M_j$-result was false. In a perfect experiment, she would win every round she chose to play; even in a practical experiment she can achieve significantly more than the 50\% success rate which would be predicted by MR.} To understand Alice's advantage we must examine the game from a QM perspective.


Alice uses the initial $M_3$ measurement to obtain the pure quantum state \ket{3}. She applies the unitary $\hat{U}_\textrm{I}$ that operates as $\hat{U}_\textrm{I} = \ketbra{I}{3} + (\textit{orthogonal terms})$, to produce the initial state
\begin{align}
\ket{I}  &= \frac{ \ket{1} + \ket{2} + \ket{3} }{\sqrt{3}}
\end{align}
Her operation presents the state \ket{I} to Bob, who next measures $M_j$ on \ket{I}, performing a projection. If Bob's $M_j$-result is true, he has applied the quantum projector $\hat{P}_j = \ketbra{j}{j}$ while by finding an $M_j$-result that is false, he has applied $\hat{P}^{\perp}_j = \mathds{1} - \ketbra{j}{j}$. \replymooji{Alice would then like to measure the component of the state left by Bob's measurement along the final state $\ket{F} =  ( \ket{1} + \ket{2} - \ket{3} ) / \sqrt{3} $. Bob's projectors on Alice's initial and final states \ket{I} and \ket{F} obey: 
\begin{align}
|\matelm{F}{\hat{P}_1}{I}|^2 &= |\matelm{F}{\hat{P}_2}{I}|^2 = 1/9 \\
|\matelm{F}{\hat{P}^\perp_1}{I}|^2 &= |\matelm{F}{\hat{P}^\perp_2}{I}|^2 = 0
\end{align}
for both $j=1$ and $j=2$. Alice cannot directly measure \ket{F}, but is able to transform  state \ket{F} into state \ket{3} with a unitary $\hat{U}_\textrm{F} = \ketbra{3}{F} + (\textit{orthogonal terms})$, and then use her measurement of $M_3$ as an effective $M_F$ measurement.} Alice therefore obtains $M_3$-true when Bob's $M_j$-result is true with probability $P_{M_3}(A \cap B) = |\matelm{3}{\hat{U}_\textrm{F} \hat{P}_j \hat{U}_\textrm{I}}{3}|^2 = |\matelm{F}{\hat{P}_j}{I}|^2 = 1/9$, and when Bob's $M_j$-result is false $P_{M_3}(A \cap  \negation{B}) = |\matelm{3}{\hat{U}_\textrm{F} \hat{P}^\perp_j \hat{U}_\textrm{I}}{3}|^2 = |\matelm{F}{\hat{P}^\perp_j}{I}|^2 = 0$. Alice finding her $M_3$-result true is conditional on Bob leaving a component of \ket{\psi_j} along \ket{F}; to do so, his $M_j$-result cannot have been false. Alice's probability conditioned on Bob is then $P_{M_j}(B | A) = 1$. Alice bets whenever  her $M_3$-result is true, playing \sfrac{1}{9} of the rounds and winning each round she plays.


\begin{figure*}[t]
\begin{center}
\includegraphics[width=\dfigwidth]{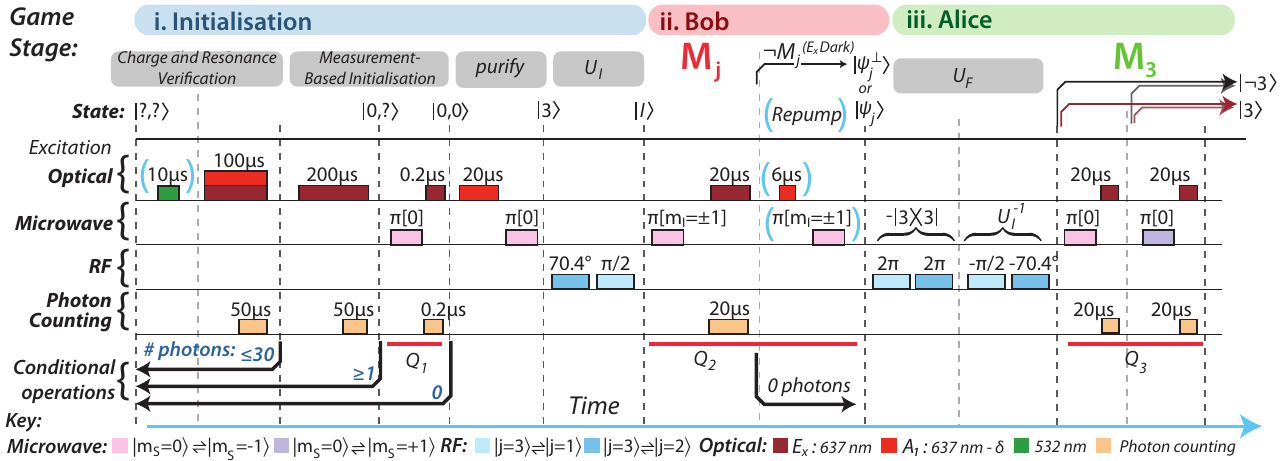}
\end{center}
\caption{
\label{fig2}
\textbf{Microwave, rf and optical pulse sequence implementing the `three-box' experiment.} 
{\it i}\sspace )   Initialisation consists of preparing the NV$^-$ state via charge-state verification (CSV) and measurement-based initialisation (MBI) into state \ket{3} followed by \johnadded{purification of $m_s=0$ and} application of $\hat{U}_I$.
  {\it ii}\sspace )~Bob's measurement $M_j$ consists of: Moving population from $m_S = -1$ to $m_S = 0$ conditioned on $m_I$, indicated $\pi [m_I ]$ in the figure, followed by monitoring of E$_x$ fluorescence. If fluorescence is observed, a `repopulating' sequence via the spectrally resolved A$_1$ fluorescence ($\lambda_{A_1} - \lambda_{E_x} = \delta = 1.89 \, \mathrm{pm}$) resets $m_S = -1$ whilst leaving $m_I$ unchanged and ready for Alice's measurement.
  {\it iii}\sspace ) Alice's measurement consists of the unitary $\hat{U}_F$, followed by read-out of $M_3$ in the $m_S = -1$ and $m_S = +1$ sublevels. Further details on the experimental sequence are provided in the \SuppMat.}
 \end{figure*}

Our implementation of this game uses the nitrogen-vacancy centre (NV$^-$), which hosts an excellent three-level quantum system for implementing this `three-box' game: the $^{14}$N nucleus which has $(2I+1) = 3$ quantum states (see Fig.~\ref{fig1}a). Although we cannot (yet) superpose a physical ball under three separate boxes, by using radio frequency (RF) pulses~\cite{Schweiger:pulse_epr_book} we can readily prepare the $^{14}$N angular momentum into a superposition of alignment along three distinct spatial axes, providing three mutually exclusive `box states' in the macrorealist picture. We work in the electron spin $m_S \!=\!-1$ manifold, and assign eigenvalues of nitrogen nuclear spin $m_I$ to the box-states $j$ according to: a) $\ket{m_I \! = \! -1 } \sim \ket{j \! = \! 1}$ b) $\ket{m_I \! = \! +1 } \sim \ket{j \! = \! 2}$ c) $\ket{m_I \! = \! 0 } \sim \ket{j \! = \! 3}$ (see Fig.~\ref{fig1}b).

Preparation and readout of the $^{14}$N nuclear spin is provided via the NV$^-$ electronic spin ($S\!=\!1$). We use selective microwave pulses to change $m_S$ conditioned on $m_I$, and then read out the electron spin in a single shot and with high fidelity~\cite{Robledo:2011fs}, by exploiting the NV centre's electron spin-selective optical transitions. The spin read-out achieves $96\%$ fidelity and takes $\approx$\,20\,$\mu$s, much shorter than the nuclear spin inhomogeneous coherence lifetime of $T_2^* \gg 1$\,ms at $T=8.7$\,K, enabling three sequential read-out operations during a single coherent evolution of the system, as required for our `three box' implementation. We achieve all steps of the quantum experiment well within the coherence time of our system and therefore make no use of refocussing pulses.

The full experimental sequence is shown in Fig.~\ref{fig2}, with further details in the \SuppMat. The initial state \ket{3} is prepared by projective nuclear spin readout utilising a short duration ($\simeq 200$\,ns) optical excitation. The subsequent experiment is then conditioned on detection of at least one photon during the preparation phase, which heralds \ket{3} with $\gtrsim 95 \%$ fidelity (Fig.~\ref{fig1}d) at the expense of $\lesssim 1\%$ preparation success rate. \ronaldremoved{\johnadded{(}} Once \ket{3} is heralded, all subsequent data is accepted unconditionally for the Leggett-Garg test.\ronaldremoved{\johnadded{)}} After initialisation, Alice transforms the state \ket{3} into \ket{I} via two RF pulses (see \SuppMat) and hands the system to Bob.

\begin{figure}[t]
\begin{center}
\includegraphics[width=\sfigwidth]{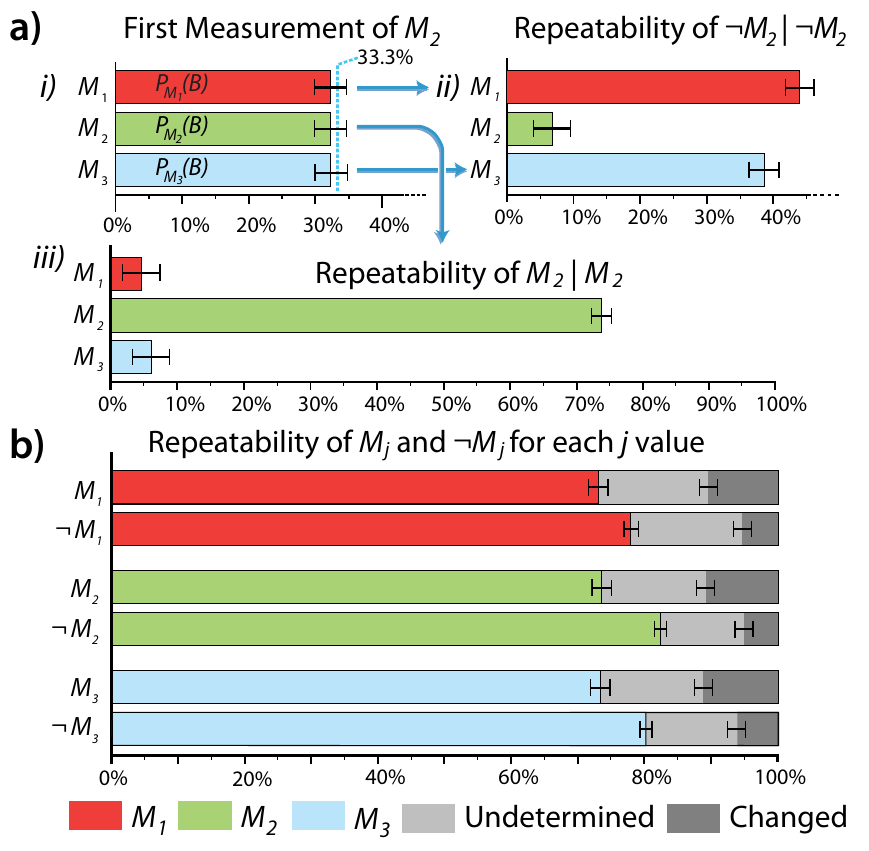}
\end{center}
\caption{\label{fig3} {\bf Bob verifies the non-detectable nature of the $M_j$ measurements.} 
{\bf a)} Measurement within the $m_S = -1$ manifold only:
i) Bob's measurement results when observing the state \ket{I} in the \ket{j} basis are independent of the $j$ value selected to within experimental error. Repeatability is illustrated by plotting the result of a second $M_j$ measurement within $m_S = -1$, conditioned on (ii) the result $\negation{M_2}$, or (iii) the result $M_2$.  {\bf b)} The repeatability of each $M_j$ measurement is studied within the $m_S = -1$ manifold; a finite probability exists for the electron spin to branch into the $m_S = +1$ manifold, yielding an undetermined reading, and for the nuclear spin to flip producing a `definitely changes' outcome. Error bars show $\pm 2 \, \sigma$ / $95\%$ confidence intervals.} 
\end{figure}

Bob picks a secret $j$-value and maps the corresponding nuclear spin projection to the electron spin by applying a microwave $\pi$-pulse to drive a transition from one of the $m_S \! = \! -1$ states (\ket{j} is \ket{1} or \ket{2}) into the $m_S \!\! = \! 0$ manifold. He then uses optical measurement of the E$_x$ fluorescence to determine $m_S$. Absence of fluorescence (an `E$_x$-dark' NV$^-$)  implies $\negation{M_j}$ and collapses the electron state into $m_S \!\! = \!\! -1$ whilst performing $\hat{P}^\perp_j$ on the nuclear spin. (Fig.~\ref{fig3}a.ii). We find that nuclear spin coherences within $m_S \!\! = \! -1$ are unaffected by the $\negation{M_j}$ read-out process. 

Detection of $n \! \ge \! 1$ photons during Bob's 20\,$\mu$s readout projects into $m_S \!\! = \! 0$ and corresponds to an $M_j$-result true. In such events, there is $\simeq \! 70\%$ chance the electronic spin will be left in an incoherent mixture of $m_S \! = \! \pm 1$ following readout, due to optical pumping~\cite{Robledo:2011fs}. Conditionally on Bob's $M_j$-result being true, we take care to undo the mixing effect of this mixing as follows: We first pump the electron spin to $m_S \! = \! 0$ by selective optical excitation of $m_S \! = \! \pm 1$ (A$_1$ light), followed by driving a selective a microwave pulse from $m_S \! = \! 0$ to $m_S \! = \! -1$ (Fig.~\ref{fig1}c). This procedure is effective, because the optical fluorescence preserves the nuclear spin populations $m_I$ that encode the game eigenstates in $\gtrsim 70\%$ of cases (see Fig.~\ref{fig3}b). Bob performs repeated pairs of measurements, verifying that from a macrorealist's perspective, that performing $M_j$ is equivalent to opening one of three boxes containing a hidden ball. Bob finds the probability for each $M_j$ is $\simeq \sfrac{1}{3}$ (Fig.~\ref{fig3}a.i). Bob performs consecutive $M_j$ observations and verifies that finding $M_j$ ($\negation{M}_j$) true on one run implies that a subsequent measurement of $M_j$ ($\negation{M}_j$) will also be true (Fig.~\ref{fig3}b,c) \addedrichard{gathering statistics over $N \! = \! 1200$ trials for each combination.}

Once Bob has measured \addedrichard{in secret}, Alice predicts his result by mapping \ket{F} to \ket{3} and performing $M_3$. Alice accomplishes this via: $\ket{F} \rightarrow \ket{I} \rightarrow \ket{3}$. The Berry's phase associated with $2 \pi$ rotations~\cite{Morton:2006iz} provides the map $\ket{F} \rightarrow \ket{I}$ via two RF pulses that change the signs of the $\left\{ \ket{1},\ket{3} \right\}$ and then $\left\{ \ket{2},\ket{3} \right\}$ states. State \ket{3} then acquires two sign changes yielding \ket{F} up to a global phase. The map $\hat{U}_I^{-1}$ from \ket{I} to \ket{3} is then achieved by inverting the order and phase of Alice's initial $\hat{U}_I$ pulses. (See \SuppMat). 

\begin{figure}[t]
\begin{center}
\includegraphics[width=\sfigwidth]{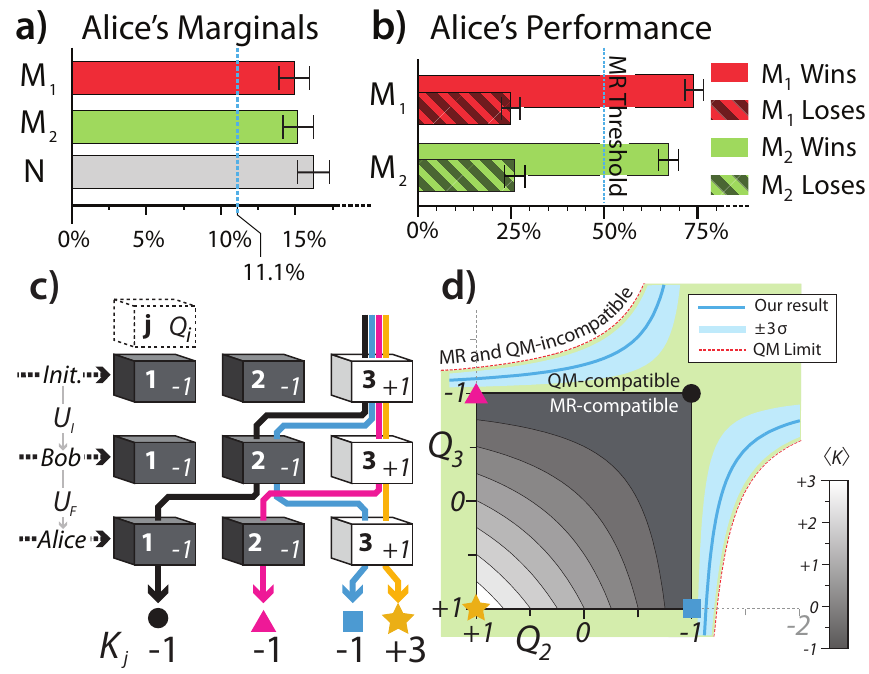} 
\caption{\label{fig4} {\bf Violation of a Leggett-Garg inequality in the `three-box' game.}
{\bf a)} Alice's measurement $M_3$ is independent of Bob's choice to perform measurement $M_1$, $M_2$ or neither ($N$). 
{ \textbf{b)} The observations of Bob and Alice are correlated to indicate the probability that Bob has (or has not) seen state $M_j$ given that Alice has seen $M_3$, determining who `wins' the game. Alice's probability to `win' exceeds 50\% for both Bob's choices $M_1$ and $M_2$. }
{\bf c)} Four MR-compatible histories that extremise $\langle K \rangle$ are illustrated by four trajectories passing through different boxes during the game. A trajectory entirely within the white $j=3$ boxes has $Q_{\{1,2,3\}} = +1$ and yields $\langle K \rangle = +3 $. Histories that visit other boxes yield $\langle K \rangle = -1$. 
{\bf d)} The $\langle K \rangle$ values of the four paths are shown the corners of the $(Q_2,Q_3)$ graph. Values for $\langle Q_2 \rangle$ and $\langle Q_3 \rangle$ from MR-compatible experiments must lie inside the shaded square satisfying $-1 \le \langle K \rangle \le 3$. Our measurement lies on the 
The value of $\langle K \rangle = -1.265$ that we measured lies cyan curve outside this region.}
\end{center}
\end{figure}

\addedrichard{Alice and Bob then compare their measurement results during $N \! = \! 2*1200$ rounds of play, distributed evenly across Bob's two choices of $M_j$ measurement. Alice finds her final $M_3$-result is true in $\simeq \!\! 15\%$ of cases, independent of Bob's choice of measurement context between $M_1$ and $M_2$ (Fig.~\ref{fig4}a). Amongst those $\simeq \!\! 15\%$ of cases where Alice's $M_3$-result is true and she chooses to bet, Bob finds she wins $\gtrsim \!\! 67\%$ of such rounds, for either of Bob's choices between measuring $M_1$ and $M_2$ (Fig.~\ref{fig4}b), confounding the macrorealist expectation. The principle source of error in our experiment arises from imperfect control of the nuclear spin.} (see \SuppMat).

To quantify the apparent incompatibility with macrorealism, we construct a Leggett-Garg function for our system, defined as:
\begin{align}
\langle K \rangle &= \langle Q_1 Q_2 \rangle + \langle Q_2 Q_3 \rangle + \langle Q_1 Q_3\rangle
\end{align}
where $Q_j$ are observables of our system recorded at three different times, derived from Alice and Bob's measurements~\cite{LEGGETT:1985td}. We assign $Q_j = +1$ whenever an $M_3$-result is true (or could be inferred true in the MR picture) and assign $Q_j=-1$ otherwise. The initially heralded state \ket{3} fixes the value of $Q_1=+1$ always, and values for $Q_2$ and $Q_3$ are taken directly from Bob and Alice's measurement results. The Leggett-Garg function satisfies $-1 \le \langle K \rangle \le +3$ for all MR systems~\cite{LEGGETT:1985td}, and for the present system, we can show $\langle K \rangle$ is related to Bob and Alice's statistics as (see \SuppMat):
\begin{align}
\langle K \rangle &= \frac{4}{9} \left( 1 - P_{M_1}( B | A ) - P_{M_2} ( B | A ) \right) - 1
\label{k_mr_expression}
\end{align}
where $P_{M_j}(B|A)$ is the probability that Bob finds the $M_j$-result true, given that Alice has also found her final $M_3$-result true. Macrorealism asserts that $M_1$ and $M_2$ are mutually exclusive events, whereas QM does not, so that:
\begin{align}
&& \mathrm{MR} & : & P_{M_1}(B|A) + P_{M_2}(B|A) &\le 1 &&  \\
&& \mathrm{QM} & : & P_{M_1}(B|A) + P_{M_2}(B|A) &\le 2 && 
\end{align}
\noindent Under QM assumptions, eqn.~\ref{k_mr_expression} satisfies $\langle K \rangle \ge -13/9 = -1.4\dot{4}$, possibly lying outside the range compatible with MR. We determined $\langle K \rangle$ by estimating $P_{M_j}(B|A)$ during $N \! = \! 1200$ trials of Bob measuring $j\!=\!1$ and $j\!=\!2$, finding $\langle K \rangle = -1.265 \pm .023$ in the quantum game, corresponding to a $\simeq \!\! 11.3 \; \sigma$ violation of the Leggett-Garg inequality under fair sampling assumptions, and $\simeq \!\! 7.8 \; \sigma$ violation in a maximally adverse macrorealist position in which all undetermined measurements are assumed to represent Alice `cheating' and are reassigned to minimise the quantum discrepancy from MR predictions (see \SuppMat).
 

Our results unite two concepts in foundational physics: Leggett-Garg inequalities~\cite{LEGGETT:1985td} and pre- and post-selected effects~\cite{AHARONOV:1964ul} in a  quantum system to which the Kochen-Specker no-go theorem applies~\cite{Kochen:1967vo}.  Experimental studies of the Leggett-Garg inequality have previously utilised ensembles~\cite{Moussa:2010ky,Knee:2011uva}, or made assumptions regarding process stationarity~\cite{Palacios-Laloy:1266756,Waldherr:2011kma}, or have required weak measurements~\cite{Goggin:2011iw} to draw conclusions, whilst the existing studies of the three box problem cannot incorporate measurement non-detectability~\cite{Resch:2004vi,Kolenderski:2011ug}, presenting a loophole that allows classical \addedrichard{non-}contextual models to reproduce the quantum statistics~\cite{Leifer:2005ek}. We have studied the `three box' experiment on a matter system, as originally conceived~\cite{AHARONOV:1991ux} and developed~\cite{Aharon:2008fv} in terms of sequential, projective non-demolition measurements, and we may therefore re-examine the conclusions that can be drawn when using this improved measurement capability.

Two assumptions underpin macrorealism; {\it 1:} macroscopic state definiteness and {\it 2:} non-disturbing measurability. In previous studies it has been possible to assign violations of the Leggett-Garg inequality to a loss of non-disturbing measurability, in both optical~\cite{Goggin:2011iw} and spin-based~\cite{Knee:2011uva} experiments. The disturbance due to measurement can sometimes be surprisingly non-local~\cite{Kwait:1997tr} and it has been suggested that detectable disturbance is a necessary condition for violating a Leggett-Garg inequality in all cases~\cite{Benatti:1994ui,FoundationsSymposium:Clifton}. Our results show explicitly that Alice cannot detect Bob's measurement (Fig.~\ref{fig4}a), so that the measurement has no detectable disturbance, whilst the statistics violate a Leggett-Garg inequality. We are therefore able to rule out the macrorealist's assumption {\it 1:} of state definiteness. 

Our experiment makes use of a three-level quantum system in which Bob's choice between $M_1$ and $M_2$ represents a choice of measurement `context' in the language of Kochen and Spekker~\cite{Kochen:1967vo}. If Bob is able to keep his measurement context secret, then a macrorealist Alice could only use a `non-contextual' classical theory to describe the experiment. It is known that every pre-and post-selection paradox implies a Kochen-Specker proof of quantum contextuality~\cite{Leifer:2005ek}.  It has been argued that measurement disturbance provides a loophole to admit non-contextuality into classical models (in addition to finite measurement precision~\cite{Meyer:1999wm,Clifton:2000ws}), all classical models presented to date that exploit this loophole give rise to detectable measurement disturbances.  In our experiment, Bob's intervening measurement introduces no disturbances detectable by Alice,  and cannot be accounted for by existing classical models.


\noindent {\bf Acknowledgements:} This work is supported by the DARPA QuEST and QuASAR programs, the Dutch Organization for Fundamental Research on Matter (FOM), the Netherlands Organization for Scientific Research (NWO) and the European Commission (DIAMANT). J.J.L.M. is supported by the Royal Society and St. John's College, Oxford. R.E.G, O.M and G.A.D.B thank the John Templeton Foundation for supporting this work.
 
\noindent {\bf Competing Financial Interests:} The authours declare no competing financial interests.

\end{document}